\documentclass{article}

\usepackage{graphicx}
\usepackage{amssymb}

\usepackage[default]{jasa_harvard}    

\usepackage{amsmath}
\DeclareMathOperator*{\argmax}{argmax} 
\usepackage{theorem}
\usepackage{latexsym}
\usepackage{bm}

\usepackage{color} 

\theorembodyfont{\itshape}
\newtheorem{Theorem}{Theorem}
\theorembodyfont{\itshape}
\newtheorem{Lemma}[Theorem]{Lemma}
\theorembodyfont{\itshape}
\newtheorem{Corollary}[Theorem]{Corollary}
\theorembodyfont{\itshape}
\newtheorem{Proposition}[Theorem]{Proposition}
\theorembodyfont{\rmfamily}
\newtheorem{Remark}{Remark}
\theorembodyfont{\rmfamily}
\newtheorem{Proof}{Proof}

\begin{document}
\title{Approximate Methods for State-Space Models}
\author{Shinsuke Koyama\thanks{Postdoctoral fellow, 
Department of Statistics and Center for the Neural Basis of Cognition, 
Carnegie Mellon University, Pittsburgh, PA 15213 
(email: koyama@stat.cmu.edu).  Corresponding author.}
\and 
Lucia Castellanos P\'erez-Bolde\thanks{Graduate Student, 
Machine Learning Department and Center for the Neural Basis of Cognition, 
Carnegie Mellon University, Pittsburgh, PA 15213 
(email: lcastell@cs.cmu.edu).}
\and
Cosma Rohilla Shalizi\thanks{Assistant Professor, 
Department of Statistics, 
Carnegie Mellon University, Pittsburgh, PA 15213 
(email: cshalizi@stat.cmu.edu), and External Faculty, Santa Fe Institute, Santa Fe, NM 87501.}
\and
Robert E. Kass\thanks{Professor, 
Department of Statistics, Machine Learning Department and Center for the Neural Basis of Cognition, 
Carnegie Mellon University, Pittsburgh, PA 15213 
(email: kass@stat.cmu.edu).}}
\date{}
\maketitle

\begin{abstract}
State-space models provide an important body of techniques for analyzing
time-series, but their use requires estimating unobserved states.  The optimal
estimate of the state is its conditional expectation given the observation
histories, and computing this expectation is hard when there are
nonlinearities.  Existing filtering methods, including sequential Monte Carlo,
tend to be either inaccurate or slow.  In this paper, we study a nonlinear
filter for nonlinear/non-Gaussian state-space models, which uses Laplace's
method, an asymptotic series expansion, to approximate the state's conditional
mean and variance, together with a Gaussian conditional distribution.  This
{\em Laplace-Gaussian filter} (LGF) gives fast, recursive, deterministic state
estimates, with an error which is set by the stochastic characteristics of the
model and is, we show, stable over time.  We illustrate the estimation ability
of the LGF by applying it to the problem of neural decoding and compare it to
sequential Monte Carlo both in simulations and with real data.  We find that
the LGF can deliver superior results in a small fraction of the computing time.

\noindent\textsc{Keywords}: {Laplace's method, recursive Bayesian estimation, neural decoding}
\end{abstract}

\section{Introduction}

The central statistical problem in applying state-space models is that of {\em
  filtering}, i.e., estimating the unobserved state from the observations.  For
nonlinear or non-Gaussian models, considerable effort has been devoted to
devising approximate solutions to the filtering problem, based mainly on
simulation methods such as particle filtering and its variants
\cite{Kitagawa87,Kitagawa96,Sequential-Monte-Carlo-book}.  In this article we
study a deterministic approximation based on sequential application of
Laplace's method which we call the {\em Laplace Gaussian filter} (LGF), and we
illustrate the approach in the context of real-time neural decoding
\cite{Brockwell-Kass-Schwartz,Eden-at-al-point-process-filtering,Serruya-Hatsopoulos-Paninski-Fellows-Donoghue}.
In this context we find the LGF to be far more accurate, for equivalent
computational cost, than particle filtering.

Suppose we have a stochastic state process
$\{x_t\},\ t=1,2,\ldots$ and a related observation process $\{y_t\}$.
Filtering
consists of estimating the state $x_t$ given a sequence of observations $y_1,
y_2, \ldots y_t \equiv y_{1:t}$, i.e., finding the posterior distribution
$p(x_t|y_{1:t})$ of the state, given the sequence.  
It is common to assume that the state $x_t$ is
a first-order homogeneous Markov process, with initial density $p(x_1)$ and
transition density $p(x_{t+1}|x_t)$, and that $y_t$ is independent of states
and observations at all other times given $x_t$, with observation density
$p(y_t|x_t)$.  Bayes's Rule gives a recursive filtering
formula,
\begin{eqnarray}
p(x_t|y_{1:t}) &=&
\frac{ p(y_t|x_t)p(x_t|y_{1:t-1})}{\int{p(y_t|x_t)p(x_t|y_{1:t-1})dx_t}},
\label{eq:filter}
\end{eqnarray}
where
\begin{equation}
p(x_t|y_{1:t-1}) = \int{p(x_t|x_{t-1})p(x_{t-1}|y_{1:t-1})dx_{t-1}}
\label{eq:predict}
\end{equation}
is the predictive distribution, which convolves the previous filtered
distribution with the transition density.  In principle, Equations
(\ref{eq:filter}) and (\ref{eq:predict}) give a complete, recursive solution to
the filtering problem for state-space models: the mean-squared optimal point
estimate is simply the mean of the posterior density given by Equation
(\ref{eq:filter}).  When the dynamics are nonlinear, non-Gaussian, or even just
high-dimensional, however, computing these estimates sequentially can be a
major challenge.

One approach to Bayesian computation is to attempt to simulate from the
posterior distribution.  Applying Monte Carlo simulation to Equations
(\ref{eq:filter})--(\ref{eq:predict}) would let us draw from $p(x_t|y_{1:t})$,
if we had $p(x_t|y_{1:t-1})$.  The insight of particle filtering 
is that the latter distribution can itself be approximated by Monte Carlo simulation
\cite{Kitagawa96,Sequential-Monte-Carlo-book}.  
This turns the recursive
equations for the filtering distribution into a stochastic dynamical system of
interacting particles \cite{Del-Moral-Miclo}, each representing one draw from
that posterior.  While particle filtering has proven itself to be useful in
practice
\cite{Sequential-Monte-Carlo-book,Brockwell-Rojas-Kass-bayesian-decoding,Ergun-Barbieri-Uri-Wilson-Brown},
like any Monte Carlo scheme it can be computationally costly; moreover, the
number of particles (and so the amount of computation) needed for a given
accuracy grows rapidly with the dimensionality of the state-space.  For
real-time processing, such as neural decoding, the computational cost of
effective particle filtering can quickly become prohibitive.

The primary difficulty with the nonlinear filtering equations comes from their
integrals.  We use Laplace's method to obtain estimates of the mean and
variance of the posterior density in Eq.\ (\ref{eq:filter}), and then
approximate that density by a Gaussian with that mean and variance.  This
distribution is then recursively updated in its turn when the next observation
is taken.

There are several versions of Laplace's method, all of which replace integrals
with series expansion around the maxima of integrands.  An expansion parameter,
$\gamma$, measures the concentration of the integrand about its peak.  In the
simplest version, the posterior distribution is replaced by a Gaussian centered
at the posterior mode.  Under mild regularity conditions, this gives a
first-order approximation of posterior expectations, with error of order
$O(\gamma^{-1})$.  Several papers have applied some form of first-order Laplace
approximation sequentially
\cite{Brown-et-al-statistical-paradigm,Eden-at-al-point-process-filtering}.  In
the ordinary static context, \citeasnoun{Tierney-Kass-Kadane} analyzed the way
a refined procedure, the ``fully exponential'' Laplace approximation, gives a
second-order approximation for posterior expectations, having an error of order
$O(\gamma^{-2})$.  In this paper we provide theoretical results justifying
these approximations in the sequential context.  Because state estimation
proceeds recursively over time, it is conceivable that the approximation error
could accumulate, which would make the approach ineffective.  Our results show
that, under reasonable regularity conditions, this does not happen: the
posterior mean from the LGF approximates the true posterior mean with error
$O(\gamma^{-\alpha})$ uniformly across time, where $\alpha=1$ or 2 depending on
the order of the LGF.

We specify the LGF in Section \ref{sec:LGF}, and give our theoretical results
in Section \ref{sec:theory}. Section \ref{sec:application} introduces the
neural decoding problem and reports comparative results both in simulation
studies and with real data.  We provide additional comments in Section
\ref{sec:discussion}.  Proofs and implementation details are collected in the appendix.\footnote{Appendices B--E appeared as a supplementary file in the journal version.}

\section{The Laplace-Gaussian filter (LGF)}
\label{sec:LGF}

Throughout the paper, $x_{t|t}$ and $v_{t|t}$ denote the mode and variance of
the true filtered distribution at time $t$ given a sequence of observations
$y_{1:t}$, and similarly $x_{t|t-1}$ and $v_{t|t-1}$ are those of the
predictive distribution at time $t$ given $y_{1:t-1}$, respectively.  Hats\
$\hat{}$\ and tildes\ $\tilde{}$\ on variables indicate approximations; in
particular, $\hat{x}$ denotes the approximated posterior mode, and $\tilde{x}$
the approximated posterior mean.  The transpose of a matrix $\bm{A}$ is written
$\bm{A}^T$.  Bold type of a small letter indicates a column vector.

\subsection{Algorithm}

The LGF procedure for a one-dimensional state is as follows. (The  
multi-dimensional extension is straightforward; see below.)

\begin{enumerate}
\item At time $t=1$, initialize the predictive distribution of the state, $p(x_1)$.
\item Observe $y_t$.
\item (Filtering) Obtain the approximate posterior mean $\tilde{x}_{t|t}$ and
  variance $\tilde{v}_{t|t}$ by Laplace's method (see below), and set
  $\hat{p}(x_t|y_{1:t})$ to be a Gaussian distribution with the same mean and
  variance.
\item (Prediction) Calculate the predictive distribution,
\begin{equation}
\hat{p}(x_{t+1}|y_{1:t}) =  \int{p(x_{t+1}|x_t)\hat{p}(x_t|y_{1:t})dx_{t}}.
\label{eq:predict_approx}
\end{equation}
\item Increment  $t$  and go to step 2.
\end{enumerate}

We consider first- and second- order Laplace's approximations.  
In the first-order Laplace approximation the posterior mean and variance are
$\tilde{x}_{t|t} = \hat{x}_{t|t} \equiv \argmax_{x_{t}}{l(x_t)}$
and
$\tilde{v}_{t|t} = [-l''(\hat{x}_{t|t})]^{-1}$, 
where
$l(x_t) = \log p(y_t|x_t)\hat{p}(x_t|y_{1:t-1})$.

The second-order (fully exponential) Laplace approximation is calculated as
follows \cite{Tierney-Kass-Kadane}.  For a given positive function $g$ of the
state, let $k(x_t)=\log{g(x_t)p(y_t|x_t)\hat{p}(x_t|y_{1:t-1})}$, and let
$\bar{x}_{t|t}$ maximize $k$.  The posterior expectation of $g$ for the second
order approximation is then
\begin{equation}
\hat{E}[g(x_t)|y_{1:t}] \approx
\frac{|-k''(\bar{x}_{t|t})|^{-\frac{1}{2}}\exp[k(\bar{x}_{t|t})]}
{|-l''(\hat{x}_{t|t})|^{-\frac{1}{2}}\exp[l(\hat{x}_{t|t})]}.
\label{eq:2nd_approx}
\end{equation}
When the $g$ we care about is not necessarily positive, a simple and practical
trick is to add a large constant $c$ to $g$ so that $g(x)+c >0$, apply Eq.\
(\ref{eq:2nd_approx}), and then subtract $c$.  The posterior mean is thus
calculated as $\tilde{x}_{t|t} = \hat{E}[x_t+c]-c$.  (In practice it suffices
that the probability of the event $\{ g(x_t)+c > 0\}$ is close to one under the
true distribution of $x_t$.  Allowing this to be merely very probable rather
than almost sure introduces additional approximation error, which however can
be made arbitrarily small simply by increasing the constant $c$. See
\citeasnoun{Tierney-Kass-Kadane} for details.)  The posterior variance is set
to be $\tilde{v}_{t|t} = [-l''(\hat{x}_{t|t})]^{-1}$, as this suffices for
second-order accuracy (see Remark \ref{rmk:laplace-approx} in Appendix
\ref{appendix:proofs}).

To use the LGF to estimate a state in $d$-dimensional space, one simply takes
the function $g$ to be each coordinate in turn, i.e., $g(\bm{x}_t)=x_{t,i}+c$
for each $i=1,2,\ldots,d$.  Each $g$ is a function of $\mathbb{R}^d \to
\mathbb{R}$, and $|-l''(\hat{x}_{t|t})|^{-\frac{1}{2}}$ and
$|-k''(\bar{x}_{t|t})|^{-\frac{1}{2}}$ in Eq.\ (\ref{eq:2nd_approx}) are
replaced by the determinants of the Hessians of $l(\hat{\bm{x}}_{t|t})$ and
$k(\bar{\bm{x}}_{t|t})$, respectively.  Thus, estimating the state with the
second-order LGF takes $d$ times as long as using the first-order LGF, since
posterior means of each component of $\bm{x}_t$ must be calculated separately.

In many applications the state process is taken to be a linear Gaussian process
(such as an autoregression or random walk) so that the integral in Eq.\
(\ref{eq:predict_approx}) is analytic.  When this integral is not done
analytically, either the asymptotic expansion (\ref{eq:approx_predict_t+1_2})
or a numerical method may be employed.  To apply our theoretical results, the
numerical error in the integration must be $O(\gamma^{-\alpha})$, where
$\gamma$ is the expansion parameter, to be discussed in section
\ref{subsec:regularityCond}, and $\alpha =1$ or $2$ depending on the order of
the LGF.

\subsection{Smoothing}
The LGF can also be used for smoothing.  That is, given the observation up to
time $T$, $y_{1:T}$, smoothed state distributions, $p(x_t|y_{1:T})$, $ t \leq
T$, can be calculated from filtered and predictive distributions by recursing
backwards \cite{Anderson-Moore-filetering}.  Instead of the true filtered and
predictive distributions, however, we now have the approximated filtered and
predictive distributions computed by the LGF.  By using these approximated
distributions, the approximated smoothed distributions can be obtained as
\begin{equation}
  \hat{p}(x_t|y_{1:T}) = 
  \hat{p}(x_t|y_{1:t}) \int\frac{\hat{p}(x_{t+1}|y_{1:T})p(x_{t+1}|x_t)}{\hat{p}(x_{t+1}|y_{1:t})}dx_{t+1}.
\label{eq:backwardapprox}
\end{equation}
We address the accuracy of LGF smoothing in Theorem \ref{thm:smoothing}.

\subsection{Implementation}
\label{sec:implementation}
Two aspects of the numerical implementation of the LGF call for special
comment: maximizing the likelihood and computing its second derivatives.  One
key point is that the Hessian in Eq.\ (\ref{eq:2nd_approx}) may be computed by
careful numerical differentiation.  Avoiding analytical derivatives saves
substantial time when fitting many alternative models.  See Appendix
\ref{appendix:second-derivs} for a brief description of our numerical procedure
for computing the Hessian matrix, and
\citeasnoun{Kass-computing-observed-information} for full details.

The log-likelihood function can be maximized by an iterative algorithm
(e.g. Newton's method), in which $\hat{x}_{t|t-1}$ and $\hat{x}_{t|t}$ would be
chosen as a reasonable starting point for maximizing $l(x_t)$ and $k(x_t)$,
respectively.  The convergence criterion also deserves some care.  Writing
$x^{(i)}$ for the value attained on the $i^{\mathrm{th}}$ step of the
iteration, the iteration should be stopped when
$\left\|x^{(i+1)}-x^{(i)}\right\| < c \gamma^{-\alpha}$, where $c$ is a
constant and $\gamma$ is the expansion parameter, to be discussed in section
\ref{par:meaningGama}, and $\alpha$ is the order of the Laplace approximation.
The value of $c$ should be smaller than that of $\gamma$ ($c=1$ is a reasonable
choice in practice).

\section{Theoretical Results}
\label{sec:theory}

For simplicity, we state the results for the one dimensional case. The
extension to the multi-dimensional case is notationally somewhat cumbersome
but conceptually straightforward.
Let $p$ and $\hat{p}$ denote the true density of a random variable and its approximation, and let $h(x_t)$ be
\begin{equation}
h(x_t) = -\frac{1}{\gamma}\log{p(y_t|x_t)p(x_t|y_{1:t-1})},
\label{eq:h_x-definision}
\end{equation}
where $\gamma$ is the expansion parameter, whose meaning will be explained later in this section.

\subsection{Regularity conditions}
\label{subsec:regularityCond}

The following properties are the regularity conditions that are sufficient for the validity of Laplace's method \cite{Erdelyi-asymptotic,Kass-Tierney-Kadane,Wojdylo}.

\begin{description}
\item[(C.1)] $h(x_t)$ is a constant-order function of $\gamma$ as $\gamma\to\infty$, 
  and is five-times differentiable with respect to $x_t$.
\item[(C.2)] 
  $h(x_t)$ has an unique  interior minimum,  and its second derivative is positive (the Hessian matrix is positive definite for multi-dimensional cases)
\item[(C.3)] $p(x_{t+1}|x_t)$ is four-times differentiable with respect to $x_t$.
\item[(C.4)] The integral
  \begin{equation}
     \int{p(x_{t+1}|x_t)\exp{[-\gamma h(x_t)]}dx_t} 
    \nonumber
  \end{equation}
  exists and is finite.
\end{description}
We also assume the following condition which prohibits ill-behaved, ``explosive'' trajectories in state space:
\begin{description}
\item[(C.5)] Derivatives of $h(x_t)$ up to fifth-order and those of $p(x_{t+1}|x_t)$ 
with respect to $x_t$ up to third-order are bounded uniformly across time.
\end{description}

Strictly speaking, $h(x_t)$ is a random variable, taking values in the space of
integrable non-negative functions of $\mathbb{R}$. This random variable is
measurable with respect to $\sigma(y_{1:t})$.  Therefore, the stated regularity
conditions only need to hold with probability 1 under the distribution of
$y_{1:t}$ \cite{Kass-Tierney-Kadane}.

In the following section we will state the theorems that ensure that, under
conditions (C.1)--(C.5), the LGF does not accumulate error over time, but first
we explain the meaning of the expansion parameter.

\paragraph{Meaning of $\gamma$}
\label{par:meaningGama}

As seen in Eq.\ (\ref{eq:h_x-definision}) and the regularity condition (C.1),
for a given state-space model, $\gamma$ is constructed by combing the model
parameters so that the log posterior density is scaled by $\gamma$ as $\gamma
\to \infty$.  In general, $\gamma$ would be interpreted in terms of sample
size, the concentration of the observation density, and the inverse of the
noise in the state dynamics; we will describe how $\gamma$ is chosen for a
neural decoding model in section \ref{sec:application}.  From the construction
of $\gamma$, the second derivative of the log posterior density, which
determines the concentration of the posterior density, is also scaled by
$\gamma$.  Therefore, the larger $\gamma$ is, the more precisely variables can
be estimated, and the more accurate Laplace's method becomes.  When the
concentration of the posterior density is not uniform across state-dimentions
in a multidimensional case, a multidimensional $\gamma$ could be taken.
Without a loss of approximation accuracy, however, a simple implementation for
this case is taking the largest $\gamma$ as a single expansion parameter.

\subsection{Main theoretical results}
\label{subsec:theorems}

\begin{Theorem}[accuracy of predictive
  distributions]\label{thm:predictive-dist}
  Under the regularity conditions (C.1)--(C.4), the $\alpha$-order LGF
  approximates the predictive distribution as
  \begin{equation}
    \hat{p}(x_t|y_{1:t-1}) = p(x_t|y_{1:t-1}) + O(\gamma^{-\beta}), 
    \nonumber
  \end{equation}
  for $t\in \mathbb{N}$, where $\beta=1$ for $\alpha=1$ and $\beta=2$ for
  $\alpha \ge 2$.  Furthermore, if condition (C.5) holds, the error term
  is bounded uniformly across time.
\end{Theorem}

The error bound can also be established for the posterior (filtered)
expectations in the following theorem.

\begin{Theorem}[accuracy of posterior expectations]\label{thm:filter}
  Under the regularity conditions (C.1)--(C.4), the $\alpha$-order LGF
  approximates the filtered conditional expectation of a  four-times differentiable 
  function $g(x)$,
  \begin{equation}
    \hat{E}[g(x_t)|y_{1:t}] = E[g(x_t)|y_{1:t}] + O(\gamma^{-\beta}),
\nonumber
  \end{equation}
  for $t\in \mathbb{N}$, with $\beta$ as in Theorem \ref{thm:predictive-dist}.
  Here $E[\cdot|y_{1:t}]$ and $\hat{E}[\cdot|y_{1:t}]$ denote the expectation
  with respect to $p(x_t|y_{1:t})$ and $\hat{p}(x_t|y_{1:t})$, respectively.
  Furthermore, if condition (C.5) holds, the error term is bounded uniformly
  across time.
\end{Theorem}

Note that the order of the error is $\gamma^{-2}$ even for $\alpha \ge 2$ both
in Theorem \ref{thm:predictive-dist} and Theorem \ref{thm:filter}.  That is,
even if higher than the second-order Laplace approximation in Step 3 of the LGF
is employed, the resulting approximation error does not go beyond the
second-order accuracy with respect to $\gamma^{-1}$.  This fact leads to the
following corollary.

\begin{Corollary}\label{corollary:2nd-order}
  The second-order approximation is the best achievable for the LGF scheme.
\end{Corollary}


The following theorem refers to stability of the LGF. It states that minor
differences in the initially-guessed distribution of the state tend to be
reduced, rather than amplified, by conditioning on further observations, even
under the Laplace's approximation.

\begin{Theorem}[stability of the algorithm]\label{thm:stability}
  Suppose that two approximated predictive distributions at time $t$ satisfy
  \begin{equation}
    \hat{p}_1(x_t|y_{1:t-1})-\hat{p}_2(x_t|y_{1:t-1}) = O(\gamma^{-\nu}),
    \nonumber
  \end{equation}
  where $\nu>0$.  Then, under the regularity conditions (C.1)-(C.4), 
  applying the LGF $u (>0)$ times leads to the difference of two
  approximated predictive distributions at time $t+u$ as
  \begin{equation}
    \hat{p}_1(x_{t+u}|y_{1:t+u-1}) -\hat{p}_2(x_{t+u}|y_{1:t+u-1}) = 
    O(\gamma^{-\nu-u}).
    \nonumber
  \end{equation}
\end{Theorem}

\begin{Theorem}\label{thm:smoothing}
  Under the regularity conditions (C.1)--(C.4), the expectation of a four-times 
  differentiable function
  $g(x)$ with respect to the approximated smoothed distribution Eq.\
  (\ref{eq:backwardapprox}) is given by
  \begin{equation}
    \hat{E}[g(x_t)|y_{1:T}] = E[g(x_t)|y_{1:T}] + O(\gamma^{-\beta}),
    \nonumber
  \end{equation}
  for $t=1,2,\ldots,T$, with $\beta$ as in Theorem \ref{thm:predictive-dist}.
  Furthermore, if condition (C.5) is satisfied, the error term is bounded
  uniformly across time.
\end{Theorem}

\subsection{Computational cost} \label{sec:computational-cost}

Assuming that the maximization of $l(x_t)$ and $k(x_t)$ is done by Newton's
method, the time complexity of the LGF goes as follows.  Let $d$ be the number
of dimensions of the state, $T$ the number of time steps, and $N$ be the sample
size.  The bottleneck of the computational cost in the first-order LGF comes
from maximization of $l(x_t)$ at each time $t$.  In each iteration of Newton's
method, evaluation of the Hessian matrix of $l(x_t)$ typically costs $O(Nd^2)$,
as $d^2$ is the time complexity for matrix manipulation.  Over $T$ time steps,
the time complexity of the first-order LGF is $O(TNd^2)$.  In the second-order
LGF, the time complexity of calculating the posterior expectation of each
$x_{t,i}$ is still $O(Nd^2)$, but calculating it for $i=1,\ldots,d$ results in
$O(Nd^3)$.  Repeating over $T$ time steps, the complexity of the second-order
LGF is $O(TNd^3)$.

For comparison, take the time complexity of a particle filter (PF) with $M$
particles.  It is not hard to check that the computational cost across time
step $T$ of the particle filter is $O(TMNd)$.  For the computational cost of
the particle filter to be comparable with an LGF, the number of particles
should be $M =O(d)$ for the first-order LGF and $M =O(d^2)$ for the
second-order LGF.

\section{Application to neural decoding}
\label{sec:application}

The problem of neural decoding consists in using an organism's neural activity
to draw inferences about the organism's environment and its interaction
therewith --- sensory stimuli, bodily states, motor behaviors, etc.\
\cite{Spikes-book}.  Scientifically, neural decoding is vital to studying
neural information processing, as reflected by the many proposed decoding
techniques \cite{Dayan-Abbott}.  Its engineering importance comes from efforts
to design brain-machine interface devices, especially neural motor prostheses
\cite{Schwartz-prosthetics} such as computer cursors, robotic arms, etc.  The
brain-machine interface devices must determine, from real-time neural
recordings, what motion the user desires the prosthesis to have.  Such
considerations have led to many proposals, emanating from
\citeasnoun{Brown-et-al-statistical-paradigm}, for neural decoding based on
state-space models \cite{Brockwell-Kass-Schwartz}.

In the rest of this section, we introduce a standard model setup for neural
decoding tasks, and identify its Laplace expansion $\gamma$.  We then simulate
the model and apply the LGF, confirming the applicability of our theoretical
results, and comparing its performance to particle filtering.  Finally, we
apply the model and the LGF to experimental data.

\subsection{Model setup}
We consider the problem of decoding a ``state process'' from the firing of an
ensemble of neurons.  Here we suppose that neurons respond to a $\bm{x}_t \in
\mathbb{R}^d$, where $d$ is the number of dimensions.  $\bm{x}_t$ may be
interpreted as two- or three-dimensional hand kinematics for motor cortical
decoding
\cite{Georgopoulos-et-al-population-coding,Kettner-Schwartz-Georgopoulos,Paninski-Fellows-Hatsopoulos-Donoghue},
or hand posture (about 20 degrees of freedom) for dexterous grasping control
\cite{Panagiotis-grasp-2007}.  We consider $N$ such neurons, and assume that
the logarithm of the firing rate of neuron $i$ is \cite{TRUC05}
\begin{equation}
\label{eqn:firing-rates}
\log \lambda_i(\bm{x}_t) = \alpha_i + \bm{\beta}_i \cdot \bm{x}_t ~.
\end{equation}
We let $y_{i,t}$ be the spike count of neuron $i$ at time-step $t$. We assume
that $y_{i,t}$ has a Poisson distribution with intensity
$\lambda_i(\bm{x}_t)\Delta$, where $\Delta$ is the duration of the short
time-intervals over which spikes are counted at each time step.  We also assume
that firing of neurons is conditionally independent with each other given
$\bm{x}_t$.  Thus the probability distribution of $\bm{y}_t$, the vector of all
the $y_{i,t}$, is the product of the firing probabilities of each neuron.  We
assume that the state model is given by
\begin{equation}
\bm{x}_t = \bm{F}\bm{x}_{t-1} + \bm{\epsilon}_t,
\label{eq:state-process}
\end{equation}
where $\bm{F} \in \mathbb{R}^{d\times d}$ and $\bm{\epsilon}_t$ is a $d$-dimensional Gaussian random variable with mean zero and covariance matrix $\bm{W} \in \mathbb{R}^{d \times d}$.

The expansion parameter $\gamma$ for this model is identified as follows.  The
second derivative of $l(x_t)=\log p(y_t|x_t)\hat{p}(x_t|y_{1:t-1})$ is
\begin{equation}
l''(\bm{x}_t) =
-\Delta\sum_{i=1}^N\bm{\beta}_i{\exp(\alpha_i + \bm{\beta}_i\cdot \bm{x}_t)
\bm{\beta}_i^T-\bm{V}_{t|t-1}^{-1}},
\nonumber
\end{equation}
where $\bm{V}_{t|t-1}$ is the covariance matrix of the predictive distribution at
time $t$.  Then, from the Cauchy-Schwarz inequality,
\begin{eqnarray}
\| l''(\bm{x}_t) \| \le 
\Delta \sum_{i=1}^N \exp(\alpha_i+\bm{\beta}_i\cdot \bm{x}_t) 
\|\bm{\beta}_i\|^2 + \|\bm{V}_{t|t-1}^{-1}\|.
\nonumber
\end{eqnarray}
Since $\|\bm{V}_{t|t-1}^{-1}\|$ is scaled by $\|\bm{W}^{-1}\|$, 
we can identify the expansion parameter:
\begin{equation}
\label{eqn:what-is-gamma} 
\gamma = \Delta\sum_{i=1}^{N}e^{\alpha_i}{\|\bm{\beta}_i\|}^2 + \|\bm{W}^{-1}\| .
\end{equation}
We see that $\gamma$ combines the number and the mean firing rate of the
neurons, the sharpness of neuronal tuning curves and the noise in the state
dynamics.

Given our assumptions, the observation model $p(\bm{y}_t|\bm{x}_t)$ and the
state transition density $p(\bm{x}_{t+1}|\bm{x}_t)$ are strictly log-concave
and have an unique interior maximum in $\bm{x}_t$, and their derivatives up to
fifth-order are uniformly bounded if the state is bounded.  Furthermore,
$h(x_t)$ is a constant-order function of $\gamma$ as $\gamma\to\infty$, which
can be seen from the construction of $\gamma$.  Thus, the regularity conditions
(C.1)--(C.5) are satisfied if the initial distribution satisfies them.


In what follows, we took the initial value for filtering to be the true state
at $t=0$.  Note that when there is no information about the initial
distribution, we could use a ``diffuse'' prior density whose covariance is
taken to be large \cite{Durbin-Koopman-2001}.  Either type of
initial condition would satisfy the regularity conditions.  We can thus
construct LGFs according to section \ref{sec:LGF}.

\subsection{Simulation study}

We performed numerical simulations to study
first and second-order LGF (labeled by LGF-1 and LGF-2, respectively)  approximations
under conditions relevant to the neural decoding problems we are  
working on.
We also compared LGF to particle filtering.
We judged performance by accuracy in computing the posterior mean
(which was determined by particle filtering with a very large number  
of particles).
However, the posterior mean contains statistical inaccuracy (due to  
limited data). We also evaluated the accuracy with which the several alternative
methods approximate the underlying true state.

\paragraph{Simulation Setup}

In each simulation run, we generated a state trajectory from a $d$-dimensional
AR process, Eq.\ (\ref{eq:state-process}), with $\bm{F}=0.94\bm{I}$ and $\bm{W}=0.019\bm{I}$, $\bm{I}$
being the identity matrix, over $T=30$ time-steps of duration $\Delta =0.03$
seconds.  We examined different number of state-dimensions, $d=6,10,20,30$.
Regardless of $d$, we observed neural activity due to the state through $N=100$ neurons,
with $\alpha_i = 2.5 + \mathcal{N}(0,1)$ and $\bm{\beta}_i$ uniformly
distributed on the unit sphere in $\mathbb{R}^d$.  Finally, the spike counts
were drawn from Poisson distributions with the firing rates
$\lambda_i(\bm{x}_t)$ given by Eq.\ (\ref{eqn:firing-rates}) above.

\paragraph{Methods}

To compare LGF state estimates to the posterior mean we first needed a
high-accuracy evaluation of the posterior mean itself.  We obtained this by
averaging results from ten independent realizations of particle filtering with
$10^{6}$ particles; the resulting approximation error in the mean integrated
squared error (MISE) is $O({10}^{-7})$, and so negligible for our purposes.  We
also applied the particle filter (PF) for comparison.  The number of particles
in the PF was chosen by consideration of computational cost; as discussed in
subsection \ref{sec:computational-cost}, a LGF-1 is comparable in time
complexity to a PF with $O(d)$ particles, and a LGF-2 is comparable to a PF
with $O(d^2)$ particles.  For the case of $d \le 30$, 100 particles (PF-100)
was about the least number at which the PF was effective and was not much more
resource-intensive than the LGF-1.  In order that the computational time of a
PF matchs that of the LGF-2, we chose 100, 300, 500 and 1000 particles for
$d=6$, 10, 20 and 30, respectively. (We label it PF-scaled.)  See also Table
\ref{table:computational-time}.

We implemented all the algorithms in Matlab, and we ran them on 
Windows computer with Pentium 4 CPU, 3.80GHz and 3.50GB of RAM.

\paragraph{Results}

\begin{figure}[htbp]
\begin{center}
\includegraphics[width=6cm]{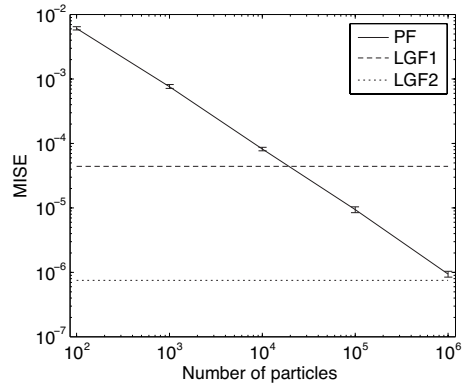}
\end{center}
\caption{Scaling of the MISE for particle filters.  The solid line represents
  the MISE (vertical axis) of the particle filter as a function of the number
  of particles (horizontal axis).  Error here is with respect to the actual
  posterior expectation (optimal filter).  The dashed and dotted horizontal
  lines represent the MISEs for the first- and second-order LGF, respectively.}
\label{fig:error-PF} 
\end{figure}

The first four rows in Table \ref{table:result1} show the four filters' MISE in
approximating the actual posterior mean.  LGF-2 gives the best approximation,
followed by LGF-1; both are better than PF-100 and PF-scaled.  Note that LGF-1
is much faster than PF-100, and the computational time of LGF-2 is
approximately the same as that of PF-scaled (Table
\ref{table:computational-time}).  Figure \ref{fig:error-PF} displays the MISE
of particle filters in approximating the actual posterior mean as a function of
the number of particles, for $d=6$.  PF needs on the order of $10^4$ particles
to be as accurate as LGF-1, and about $10^6$ particles to match LGF-2.
Furthermore, since the computational time of the PF is proportional to the
number of particles, the time needed to decode by PF with $10^4$ and $10^6$
particles are expected to be about 20s and 2,000s, respectively (from Table
\ref{table:computational-time}).  Thus, if we allow the LGFs and the PF to have
the same accuracy, LGF-1 is about $1,000$ times faster than the PF, and LGF-2
is expected to be about $10,000$ times faster than the PF.

\begin{table}[htbp]
  \caption{MISEs for different filters}
  \begin{center}
    \begin{tabular}{ccccc}
      \hline\hline
      \multicolumn{1}{c}{Method}
      & \multicolumn{4}{c}{Number of dimensions, $d$} \\
      & 6 & 10 & 20  & 30\\
      \hline
      LGF-2   & 0.0000008 &   0.000002 &   0.00001  &  0.00006 \\
      LGF-1   &  0.00003 &   0.00004  &  0.0001 &   0.0002 \\
      PF-100  & 0.006 &  0.01 &  0.03 &  0.04 \\ 
       PF-scaled &  0.006  &  0.007 &  0.01 &  0.02 \\   \hline
      posterior & 0.03 &  0.04 &  0.06  & 0.07 \\ \hline
    \end{tabular}
  \end{center}
    \label{table:result1}
      NOTE: The first four rows give the
      discrepancy between four approximate filters and the optimal filter
      (approximation error).  
      The fifth row gives the MISE between
      the true state and the estimate of the optimal filter, i.e., the actual
      posterior mean (statistical error).  
      All values are means from 10 independent replicates.
      The simulation standard errors are all smaller than the leading 
      digits in the table.
\end{table}

\begin{table}[htbp]
  \caption{Time (seconds) needed to decode}
  \begin{center}
    \begin{tabular}{ccccc}
      \hline\hline
      \multicolumn{1}{c}{Method}
      & \multicolumn{4}{c}{Number of dimensions, $d$} \\
      & 6 & 10 & 20  & 30\\
      \hline
      LGF-2   & 0.24 &   0.43 & 1.0  &  2.0 \\
      LGF-1   &  0.018 &  0.024 &  0.032 &   0.056 \\
      PF-100  & 0.18 &  0.18 &  0.18 &  0.19 \\ 
       PF-scaled &  0.18  & 0.50 &  0.81 &  1.8 \\   \hline
    \end{tabular}
  \end{center}
    \label{table:computational-time}
      NOTE: All values are means from 10 independent replicates.
      The simulation standard errors are all smaller than the leading 
      digits in the table.
\end{table}

The value of $\gamma$ for this state-space model is $\gamma \approx 100$
(Eq. (\ref{eqn:what-is-gamma})).  
From Theorem \ref{thm:filter}, the MISEs of LGF-1 and LGF-2 are,
respectively, evaluated as $c_1^2\gamma^{-2}$ and $c_2^2\gamma^{-4}$, where
$c_1$ and $c_2$ are constants depending on the model parameters.  
If $c_1$ and $c_2$ were in the range 1 to 10, then  
the MISEs of LGF-1 and LGF-2 should be 
$10^{-4}$ to $10^{-6}$,  
roughly matching the simulation results.

The fifth row of Table \ref{table:result1} shows the MISE between the true
state and the actual posterior mean.  The error in using the optimal filter,
i.e., the actual posterior mean, to estimate the true state is statistical
error, inherent in the system's stochastic characteristics, and not due to the
approximations.  The statistical error is an order of magnitude larger than the
approximation error in the LGFs, so that increasing the accuracy with which the
posterior expectation is approximated does little to improve the estimation of
the state.  The approximation error in the PFs, however, becomes on the same
order as the statistical error when the state dimension is larger ($d=20$ or
30).  In such cases the inaccuracy of the PF will produce comparatively
inaccurate estimates of the true state.

\begin{figure}[htbp]
\begin{center}
\includegraphics[width=6cm]{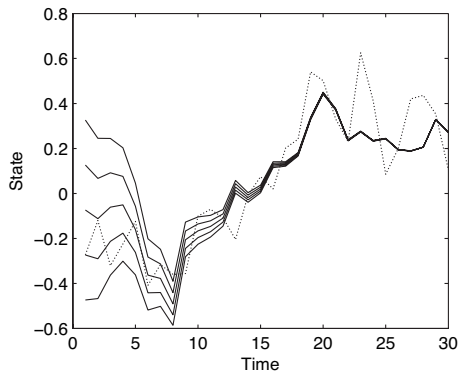}
\end{center}
\caption{The solid lines represent the estimated trajectories with five
  different initial values by LGF-1. The dashed line represents the true state
  trajectory.  }
\label{fig:insensitivity} 
\end{figure}

Finally, we examined how the choice of initial prior density affects the
filtering result.  Figure \ref{fig:insensitivity} shows five estimated
trajectories started with different initial values.  These five trajectories
converged to the same state as the time evolves, as expected from Theorem
\ref{thm:stability}.

\subsection{Real data analysis}

\paragraph{Experiment setting and data collection}

We used LGF to estimate the hand motion from neural activity.  A
multi-electrode array was implanted in the motor cortex of a monkey to record
neural activity following procedures similar to those described previously in
\citeasnoun{Velliste-Cortical-Control-Nature-2008}.  In all, 78 distinct
neurons were recorded simultaneously. Raw voltage waveforms were thresholded
and spikes were sorted to isolate the activity of individual cells.  A monkey
in this experiment was presented with a virtual 3-D space, containing a cursor
which was controlled by the subject's hand position, and eight possible targets
which were located on the corners of a cube.  The task was to move the cursor
to a highlighted target from the middle of the cube; the monkey received a
reward upon successful completion.  In our data each trial consisted of time
series of spike-counts from the recorded neurons, along with the recorded hand
positions, and hand velocities found by taking differences in hand position at
successive $\Delta = 0.03$s intervals.  Each trial contained 23 time-steps on
average. Our data set consisted of 104 such trials.

\paragraph{Methods}
For decoding, we used the same state-space model as in our simulation study.
Many neurons in the motor cortex fire preferentially in response to the
velocity $\bm{v}_t \in \mathbb{R}^3$ and the position $\bm{z}_t \in
\mathbb{R}^3$ of the hand \cite{Wang-Chan-Heldman-Moran}. 
We thus took the state  $\bm{x}_t$ to be a 6-dimensional concatenated vector 
$\bm{x}_t=(\bm{z}_t,\bm{v}_t)$. 
The state model was taken to be
\begin{equation}
 \bm{x}_t = 
 \Biggl( \begin{array}{cc}
      \bm{I} & \Delta \bm{I} \\
      \bm{O} & \bm{I}
    \end{array} \Biggr)
 \bm{x}_{t-1} + 
  \Biggl( 
  \begin{array}{c}
      \bm{0} \\
      \bm{\epsilon}_t
    \end{array} 
    \Biggr),
\end{equation}
where $\bm{\epsilon}_t$ is a 3-D Gaussian random variable with mean zero and
covariance matrix $\sigma^2 \bm{I}$, $\bm{I}$ being the identity matrix.  16 trials
consisting of 2 presentations of each of the 8 targets, were reserved for
estimating the parameters of the model.  The parameters in the firing rate,
$\alpha_i$ and $\bm{\beta}_i$, were estimated by Poisson regression of spike
counts on cursor position and velocity, and the value of $\sigma^2$ was
determined via maximum likelihood.  The time-lag between the hand movement and
each neural activity was also estimated from the same training data. This was
done by fitting a model over different values of time-lag ranging from 0 to
$3\Delta$s.  The estimated optimal time-lag was the value at which the model
had the highest $R^2$.  Having estimated all the parameters, cursor motions
were reconstructed from spike trains for the other 88 trials, and it is on
these trials we focused.  For comparison, we also reconstructed the cursor
motion with a PF-100 and a widely-used population vector algorithm (PVA)
\cite[pp.\ 97--101]{Dayan-Abbott} (see also Appendix \ref{appendix:pva}).

\begin{figure}[htbp]
\begin{center}
\includegraphics[width=13cm]{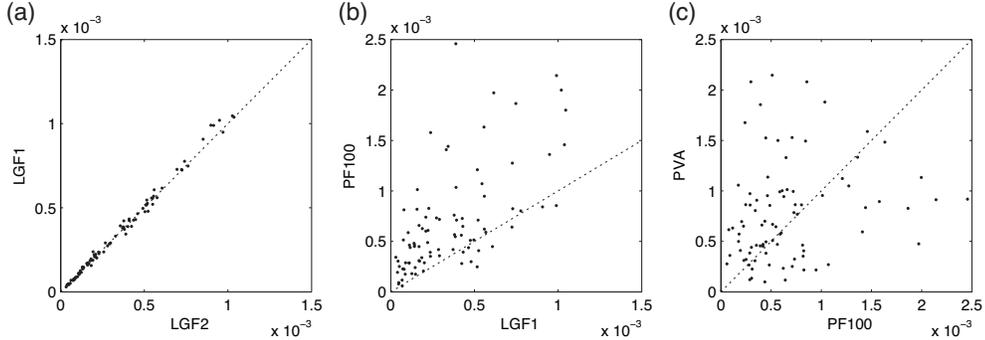}
\end{center}
\caption{Algorithm comparisons.  The horizontal and vertical axes represent the
  MISE of different algorithms in estimating the true cursor position.  Each
  point compares two different algorithms for a trial.  Overall, 4 algorithms
  (LGF-1, LGF-2, PF-100 and PVA) were compared for 88 trials.  (a) LGF-2
  vs. LGF-1, (b) LGF-1 vs. PF-100, and (c) PF-100 vs. PVA.}
\label{fig:real-data} 
\end{figure}

\paragraph{Results}
Figure \ref{fig:real-data} compares MISEs for different algorithms in
estimating the true cursor position.  Figure \ref{fig:real-data} (a) compares
the MISE of LGF-1 with that of LGF-2.  Just like in the simulation study, there
is no substantial difference between them since the statistical error is larger
than the LGFs' approximation errors.  Figure \ref{fig:real-data} (b) compares
LGF-1 to PF-100: the former estimates the true cursor position better than the
latter in most trials.  Also (Table \ref{table:computational-time}), LGF-1 is
much faster than PF-100.  Figure \ref{fig:real-data} (c) shows that the
numerical error in the PF-100 is of the same order as the error resulting from
using PVA.  (Plots of the true and reconstructed cursor trajectories are shown
in Appendix \ref{sec:real-data-trajectory}.)

\section{Discussion}
\label{sec:discussion}

In this paper we have shown that, under suitable regularity conditions, the
error of the LGF does not accumulate across time.  In the context of a neural
decoding example we found the LGF to be much more accurate than the particle
filter with the same computational cost: in our simulation study the
first-order and second-order LGFs had MISE of about 1/200 to 1/7,500 the size
of the particle filter.  We also found that for 6-dimensional case, about
10,000 particles were required in order for the particle filtering to become
competitive with the first-order LGF; and the second-order LGF remained as
accurate as the particle filter with 1,000,000 particles.  In many situations
(such as some neural decoding applications), implementation needs to be easy so
that repeated refinements in modeling assumptions may be carried out quickly.
With this in mind, it might be argued that the simplicity of the particle
filter gives it some advantages. We have, however, noted how numerical methods
may be used to supply the necessary second-derivative matrices (see Appendix
\ref{appendix:second-derivs}, and
\citeasnoun{Kass-computing-observed-information}), and these, together with
maximization algorithms, make it as easy to modify the LGF for new variations
on models as it is to modify the particle filter.  Nor does the use of the LGF
interfere with diagnostic tests and model-adequacy checks, such as the
time-rescaling theorem for point processes \cite{Brown-et-al-time-rescaling}.
The obvious conclusion is that the LGF is likely to be preferable to the
particle filter in applications where the posterior in Eq.\ (\ref{eq:filter})
becomes concentrated.

We should note that the validity of the LGF is guaranteed only when the
posterior distribution is uni-modal and has a log-concave property.  On the
other hand, the particle filter is a distribution-free method and can be used
in a multi-modal case.

It is perhaps worth emphasizing the distinction between the LGF and other
alternatives to the Kalman filter.  The simplest non-linear filter, the {\em
  extended Kalman filter} (EKF) \cite{Ahmed-filtering}, linearizes the state
dynamics and the observation function around the current state estimate
$\hat{x}$, assuming Gaussian distributions for both.  The error thus depends on
the strength of the quadratic nonlinearities {\em and} the accuracy of
preceding estimates, and so error can accumulate dramatically.  The LGF makes
no linear approximations---every filtering step is a (generally simple)
nonlinear optimization---nor does it need to approximate either the state
dynamics or the observation noise as Gaussians.

In our simulation studies, the second-order LGF was always more (in some cases
much more) than 20 times as accurate as the first-order LGF in approximating
the posterior, but this translated into only small gains in decoding
accuracy. The reason is simply that the inherent statistical error of the
posterior itself was much larger than the numerical error of the first-order
LGF in approximating the posterior.  We would expect this to be the case quite
generally.  Thus, our work may be seen as supporting the use of the first-order
LGF, as applied to neural decoding in
\citeasnoun{Brown-et-al-statistical-paradigm}.

Finally, an interesting idea is to use a sequential approximation 
to the posterior based on some
well-behaved and low-dimensional parametric family, and to apply
sequential simulation based on that family. The Gaussian 
could again be used (e.g.,
\cite{Azimi-Sadjadi-Krishnaprasad,Borigo-Hanzon-LeGland-projection,Ergun-Barbieri-Uri-Wilson-Brown}), and our results would provide
new theoretical justification for such procedures. However, it is
well-known that Gaussian distributions, with their very thin tails, 
are poorly suited for importance sampling, so that heavier-tailed
alternatives often work better (e.g., \cite{Evans-Swartz-Statstical-Science}). 
Sequential simulation schemes 
with approximating Gaussians replaced
by multivariate $t$, or other heavy-tailed distributions, may be
worth exploring.

\section*{Acknowledgment}
This work was supported by grants
RO1 MH064537,
RO1 EB005847 and  
RO1 NS050256.


\newpage

\appendix

\makeatletter   
 \renewcommand{\@seccntformat}[1]{APPENDIX~{\csname the#1\endcsname}.\hspace*{1em}}
 \makeatother

\section{Proofs of theorems} \label{appendix:proofs}

We begin by proving a lemma and a proposition needed for the main theorems.  To
simplify notation we introduce the symbols $h_t^{(l)}
\equiv \partial^lh(x_t)/\partial x_t^l |_{x_t=x_{t|t}}$ and
$q^{(l)}(x_{t+1})\equiv \partial^lp(x_{t+1}|x_t)/\partial x_t^l
|_{x_t=x_{t|t}}$.
\begin{Lemma}\label{lem:approx_mean_var}
  Let $\hat{h}(x_t)$ be
  \begin{equation}
    \hat{h}(x_t) = -\frac{1}{\gamma}\log{p(y_t|x_t)\hat{p}(x_t|y_{1:t-1})} ~,
    \label{eq:hat_h_def}
  \end{equation}
  $\hat{h}_t^{(l)} \equiv \partial^l_{x_t}\hat{h}(\hat{x}_{t|t})$, and
  $\hat{x}_{t|t}$ the minimizer of $\hat{h}(x_t)$.  Then, under the regularity
  conditions, the order-$\alpha$ Laplace approximation of the posterior mean
  and variance have series expansions as
  \begin{equation}
    \tilde{x}_{t|t} = \sum_{j=0}^{\alpha-1}{A_j(\{\hat{h}_t^{(l)}\}) \gamma^{-j}} ~,
    \label{eq:mean_approx} 
  \end{equation}
  and
  \begin{equation}
    \tilde{v}_{t|t} = 
    \sum_{j=1}^{\alpha-1}{B_j(\{\hat{h}_t^{(l)}\}) \gamma^{-j}} ~,
    \label{eq:var_approx}
  \end{equation}
  where the coefficients, $A_j$ and $B_j$, are functions of
  $\{\hat{h}_t^{(l)}\}$.
\end{Lemma}
\begin{Proof}[Lemma \ref{lem:approx_mean_var}] The expectation of a function
  $g(x_t)$ with respect to the approximated posterior distribution is
  \begin{equation}
    \hat{E}[g(x_t)|y_{1:t}] =
  \frac{\int{g(x_t)\exp{[-\gamma\hat{h}(x_t)]}dx_t}}{\int{\exp{[-\gamma\hat{h}(x_t)]}dx_t}}~,
    \label{eqn:approx-posterior-mean}
  \end{equation}
  where $g(x_t)=x_t$ for the mean and $g(x_t)=x_t^2$ for the second moment.  We
  get the coefficients $A_j$ and $B_j$ by applying Laplace's method, an
  (infinite) asymptotic expansion of a Laplace-type integral (Theorem 1.1 in
  \cite{Wojdylo}; see Appendix \ref{app:laplace} for a brief summary), to both
  the numerator and the denominator of Eq.\ (\ref{eqn:approx-posterior-mean});
  those formulae also show that the coefficients are functions of
  $\{\hat{h}_t^{(l)}\}$, $l = 1,2,\ldots$..  For example, the coefficients of
  up to first-order terms are obtained as $A_0(\{\hat{h}_t^{(l)}\}) =
  \hat{x}_{t|t}$, $A_1(\{\hat{h}_t^{(l)}\}) = -
  \hat{h}_t'''/(2{(\hat{h}_t'')}^2)$, and $ B_1(\{\hat{h}_t^{(l)}\}) =
  (\hat{h}_t'')^{-1}$.  \hfill $\Box$
\end{Proof}
\begin{Remark}\label{rmk:laplace-approx}
  Lemma \ref{lem:approx_mean_var} guarantees that the choice of
  $\tilde{x}_{t|t}=\hat{x}_{t|t}$ and
  $\tilde{v}_{t|t}=(\gamma\hat{h}_t'')^{-1}$ provides the first-order
  approximation of posterior mean and variance.  As proved in
  \citeasnoun{Tierney-Kass-Kadane}, Eq.\ (\ref{eq:2nd_approx}) achieves the
  second-order expansion of the posterior mean $\tilde{x}_{t|t}=\hat{x}_{t|t}
  +A_1(\{\hat{h}_t^{(l)}\})\gamma^{-1}$.  Thus Eq.\ (\ref{eq:2nd_approx}) and
  $\tilde{v}_{t|t}=(\gamma\hat{h}_t'')^{-1}$ provide the second-order
  approximation.
\end{Remark}

\begin{Proposition}\label{thm:predict}
  Suppose that the regularity conditions (C.1)--(C.4) hold, and that the
  approximated predictive distribution of time $t$ satisfies
  \begin{equation}
    \hat{p}(x_t|y_{1:t-1}) =
    p(x_t|y_{1:t-1}) + \sum_{j=\nu}^{N} \mathcal{E}_{t,j}(x_t)\gamma^{-j} 
    + O(\gamma^{-N-1}),
    \label{eq:approx_predict_t}
  \end{equation}
  where $\mathcal{E}_{t,j}(x_t)$ is a constant-order function of $\gamma$ and
  $0 < \nu < N$ for $\nu,N\in\mathbb{N}$.  Replacing the filtered distribution
  at time $t$ with a Gaussian with $\alpha$-order Laplace approximated mean and
  variance leads to the approximate predictive distribution at time $t+1$,
  \begin{equation}
    \hat{p}(x_{t+1}|y_{1:t}) = p(x_{t+1}|y_{1:t}) + 
    \sum_{j=\beta}^N \mathcal{E}^*_{t+1,j}(x_{t+1})\gamma^{-j} 
    + \sum_{j=\nu}^N \mathcal{E}_{t+1,j+1}(x_{t+1})\gamma^{-j-1}  + O(\gamma^{-N-1}), 
    \label{eq:approx_predict_t+1}
  \end{equation}
  where $\beta=1$ for $\alpha=1$ and $\beta=2$ for $\alpha \ge 2$.  Here
  $\mathcal{E}^*_{t+1,j}(x_{t+1})$ does not depend on 
  $\{\mathcal{E}_{t,k}(x_t)\}_{k=\nu,\nu+1,\ldots}$
  and
  \begin{equation}
    \mathcal{E}_{t+1,j+1}(x_{t+1}) =
    \frac{q'(x_{t+1})}{h_t''}\frac{\partial}{\partial x_t}
    \biggl( \frac{\mathcal{E}_{t,j}(x_t)}{p(x_t|y_{1:t-1})} \biggr)
    \Bigg|_{x_t=x_{t|t}} + O(\gamma^{-1}),
    \label{eq:propagated_error}
  \end{equation}
  for $j=\nu, \nu+1,\ldots,N$.  Furthermore, if the condition (C.5) is
  satisfied, the coefficients of the expansion terms in Eq.\
  (\ref{eq:approx_predict_t+1}) are bounded uniformly across time.
\end{Proposition}

\begin{Proof}[Proposition \ref{thm:predict}] The proof works by comparing the
  asymptotic expansions of the true and approximated predictive distributions.
  To do this, we must find those asymptotic expansions; once this is done the
  remaining steps are fairly straightforward.

  (i) We begin by evaluating the true predictive distribution at time $t+1$.
  From Eqs.\ (\ref{eq:filter}) and (\ref{eq:predict}), this is
  \begin{equation}
    p(x_{t+1}|y_{1:t}) =
    \frac{\int{p(x_{t+1}|x_t)\exp{[-\gamma h(x_t)]}dx_t}}{\int{\exp{[-\gamma h(x_t)]}dx_t}} ~.
 \nonumber
  \end{equation}
  Applying Laplace's method (Theorem 1.1 in \cite{Wojdylo}, see also Appendix
  \ref{app:laplace}) to both the numerator and the denominator of above
  equation leads to
  \begin{eqnarray}
    p(x_{t+1}|y_{1:t}) &=& 
    \frac{\sum_{s=0}^N{\Gamma(s+\frac{1}{2})(\frac{2}{h_t^{\prime\prime}})^sc^*_{2s}\gamma^{-s}+O(\gamma^{-N-1})}}{\sum_{s=0}^N{\Gamma(s+\frac{1}{2})(\frac{2}{h_t^{\prime\prime}})^s\bar{c}^*_{2s}\gamma^{-s}}+O(\gamma^{-N-1})} ~,
    \label{eq:predictive-t+1-expansion}
  \end{eqnarray}
  where
  \begin{equation}
    c^*_s = \sum_{i=0}^s{\frac{q^{s-i}(x_{t+1})}{(s-i)!}
    \sum_{j=0}^i{\Biggl(\begin{array}{c}
      -\frac{s+1}{2} \\  j
    \end{array}
    \Biggr)
    \mathcal{C}_{i,j}(A_1,\ldots)}} ~,
  \end{equation}
  and
  \begin{equation}
    \bar{c}^*_s = 
    \sum_{j=0}^s{\Biggl(
    \begin{array}{c}
      -\frac{s+1}{2} \\  j
    \end{array}
    \Biggr)
    \mathcal{C}_{s,j}(A_1,\ldots)} ~.
    \label{eq:coefficient-of-predictive-t+1}
  \end{equation}
  Here $\mathcal{C}_{s,j}(A_1,\ldots)$ is a partial ordinary Bell polynomial,
  which is the coefficient of $x^i$ in the formal expansion of
  $(A_1x+A_2x^2+\cdots)^j$, and $A_i \equiv A_i(\{h_t^{(l)}\})$ is the
  coefficient which appeared in Lemma~\ref{lem:approx_mean_var}.  Expanding
  with respect to $\gamma^{-1}$, we obtain the asymptotic expansion of
  $p(x_{t+1}|y_{1:t})$ as
  \begin{equation}
    p(x_{t+1}|y_{1:t})
    =
    q(x_{t+1}) + \sum_{j=1}^N{C_j(x_{t+1})\gamma^{-j}} + O(\gamma^{-N-1}) ~,
    \label{eq:true_predict}
  \end{equation}
  where $q(x_{t+1})$ was earlier defined as $p(x_{t+1}|x_t)$, and where
  $C_j(x_{t+1})$ depends on $q^{(k)}(x_{t+1})$ and $h_t^{(l)}\
  (k,l=1,2,\ldots)$.  $C_j(x_{t+1})$ is directly calculated by Eqs.\
  (\ref{eq:predictive-t+1-expansion})--(\ref{eq:coefficient-of-predictive-t+1}).

  (ii) We next consider the approximated predictive distribution of time $t+1$,
  \begin{equation}
    \hat{p}(x_{t+1}|y_{1:t}) =  \int{p(x_{t+1}|x_t)\hat{p}(x_t|y_{1:t})dx_t} ~,
    \label{eq:approx_predict_t+1_1}
  \end{equation}
  where $\hat{p}(x_t|y_{1:t})$ is the Gaussian distribution whose mean and
  variance are given by Eq.\ (\ref{eq:mean_approx}) and (\ref{eq:var_approx}),
  respectively.  Eq.\ (\ref{eq:approx_predict_t+1_1}) can be re-written as
  \begin{equation}
    \hat{p}(x_{t+1}|y_{1:t}) =  \frac{1}{\sqrt{2\pi\tilde{v}_{t|t}}} \int{p(x_{t+1}|x_t)
      \exp{\biggl[
          -\frac{(x_t-\tilde{x}_{t|t})^2}{2\tilde{v}_{t|t}}
        \biggr]} dx_t}.
        \nonumber
  \end{equation}
  Applying Laplace's method again,
  \begin{equation}
    \hat{p}(x_{t+1}|y_{1:t}) = \tilde{q}(x_{t+1}) +
    \sum_{j=1}^N{\frac{\tilde{q}^{(2j)}(x_{t+1})}{2^j\Gamma(j+1)}\tilde{v}_{t|t}^j}
    +O(\tilde{v}_{t|t}^{-N-1}),
    \label{eq:approx_predict_t+1_2}
  \end{equation}
  where $\Gamma(j+1)$ is the Gamma function and
  \begin{equation}
    \tilde{q}_t^{(l)} \equiv \frac{\partial^l p(x_{t+1}|x_t)}{\partial x_t}\bigg|_{x_t=\tilde{x}_{t|t}}.
    \label{eq:tilde_q}
  \end{equation}

  Now we compare Eqs.\ (\ref{eq:approx_predict_t+1_2}) and
  (\ref{eq:true_predict}), via a series of substitutions.  We want to re-write
  Eq.\ (\ref{eq:approx_predict_t+1_2}) with $q^{(k)}(x_{t+1})$ and $h_t^{(l)}$.
  Substituting Eq.\ (\ref{eq:approx_predict_t}) into Eq.\ (\ref{eq:hat_h_def}),
  \begin{eqnarray}
    \hat{h}(x_t) &=&  -\frac{1}{\gamma}
    \log p(y_t|x_t)\biggl[
      p(x_t|y_{1:t-1}) +  \sum_{j=\nu}^N \mathcal{E}_{t,j}(x_t)\gamma^{-j} + O(\gamma^{-N-1})
      \label{eq:hat_h}
    \biggr] \nonumber\\
    &=& h(x_t) - \sum_{j=\nu}^N \mathcal{F}_{t,j}(x_t)\gamma^{-j-1} + O(\gamma^{-N-2}),
    \label{eq:hat_h_h}
  \end{eqnarray}
  where
  \begin{equation}
    \mathcal{F}_{t,j}(x_t) = \frac{\mathcal{E}_{t,j}(x_t)}{p(x_t|y_{1:t-1})} + O(\gamma^{-\nu}). \nonumber
  \end{equation}
  is a collection of terms which depend on $\mathcal{E}_{t,j}(x_t)$.

  Suppose $\hat{x}_{t|t}=x_{t|t}+\epsilon$ and $\epsilon \ll 1$.  Taking the
  derivative both sides of Eq.\ (\ref{eq:hat_h_h}) and evaluating it at
  $x_{t|t}$, we obtain
  \begin{equation}
    \epsilon = \sum_{j=\nu}^N\frac{\mathcal{F}'_{t,j}}{h_t''}\gamma^{-j-1}+O(\gamma^{-N-2}). \nonumber
  \end{equation}
  Then we get
  \begin{equation}
    \hat{x}_{t|t}=x_{t|t} + \sum_{j=\nu}^N\frac{\mathcal{F}'_{t,j}}{h_t''}\gamma^{-j-1} + O(\gamma^{-N-2}).
    \label{eq:hat_x}
  \end{equation}

  Inserting Eq.\ (\ref{eq:hat_x}) into Eq.\ (\ref{eq:hat_h}) gives
  \begin{equation}
    \hat{h}_t^{(l)} = h_t^{(l)} -\sum_{j=\nu}^N \biggl[ \mathcal{F}_{t,j}^{(l)} - \frac{\mathcal{F}_{t,j}'h_t^{(l+1)}}{h_t''} \biggr]\gamma^{-j-1} + O(\gamma^{-N-2}).
    \label{eq:hat_dh}
  \end{equation}

  Substituting Eq.\ (\ref{eq:hat_x}) and Eq.\ (\ref{eq:hat_dh}) into Eq.\
  (\ref{eq:mean_approx}) leads to
  \begin{eqnarray}
    \tilde{x}_{t|t} &=&
    x_{t|t} +  \sum_{j=1}^{\alpha-1}A_j\gamma^{-j}
    + \sum_{j=\nu}^N\frac{\mathcal{F}_{t,j}'}{h_t''}\gamma^{-j-1} + O(\gamma^{-N-2}).
    \label{eq:tilde_x}
  \end{eqnarray}

  Inserting Eq.\ (\ref{eq:tilde_x}) into Eq.\ (\ref{eq:tilde_q}) and expanding
  with respect to $\gamma^{-1}$,
  \begin{eqnarray}
    \tilde{q}^{(l)}(x_{t+1}) &=& 
    q^{(l)}(x_{t+1}) 
    + \sum_{j=1}^{\alpha-1} A_j q^{(l+1)}(x_{t+1})\gamma^{-j} 
    + \sum_{j=2}^{\alpha} 
    \biggl[ \sum_{k=2}^j\frac{1}{k!}q^{(l+k)}(x_{t+1})\mathcal{C}_{j,k}(A_1,\ldots)\biggr]
    \gamma^{-j} \nonumber\\
    & & { } + 
    \sum_{j=\nu}^N\frac{\mathcal{F}_{t,j}'}{h_t''}q^{(l+1)}(x_{t+1})\gamma^{-j-1}
    +O(\gamma^{-\alpha-1}).
    \label{eq:tilde_q_2}
  \end{eqnarray}

  Substituting Eqs.\ (\ref{eq:var_approx}), (\ref{eq:hat_dh}) and
  (\ref{eq:tilde_q_2}) into Eq.\ (\ref{eq:approx_predict_t+1_2}), we obtain the
  final asymptotic expansion of $\hat{p}(x_{t+1}|y_{1:t})$,
  \begin{eqnarray} 
    \label{eq:approx_predict}   \lefteqn{\hat{p}(x_{t+1}|y_{1:t})=} & &\\
    \nonumber& &    q(x_{t+1}) + \sum_{j=1}^{\alpha} R_j(x_{t+1})\gamma^{-j} + 
    \sum_{j=\nu}^{N-1}\frac{\mathcal{F}_{t,j}'}{h_t''}q'(x_{t+1})\gamma^{-j-1}
    + O(\gamma^{-\alpha-1}) ~,
  \end{eqnarray}
  in which
  \begin{equation}
    R_j(x_{t+1}) = \left\{
      \begin{array}{ll}
        G_j(x_{t+1}) + A_jq'(x_{t+1}) & 1 \le j \le \alpha-1 \\
        G_j(x_{t+1}) & j=\alpha
      \end{array}
    \right. \nonumber
  \end{equation}
  and
  \begin{eqnarray}
    G_j(x_{t+1}) &=&
    \sum_{s=2}^j\frac{1}{s!} \mathcal{C}_{j,s}(A_1,\ldots) q^{(s)}(x_{t+1}) 
    + \sum_{s=1}^j \frac{\mathcal{C}_{j,s}(B_1,\ldots)q^{(2s)}(x_{t+1})}{2^s\Gamma(s+1)}
    \nonumber\\
    & & { } +
    \sum_{s=1}^{j-1}\sum_{k=s}^{j-1} \frac{A_{j-k}\mathcal{C}_{k,s}(B_1,\ldots)q^{(2s+1)}(x_{t+1})}{2^s\Gamma(s+1)} \nonumber\\
    & & { } + 
    \sum_{s=1}^{j-2}\sum_{k=s}^{j-2}\sum_{n=2}^{j-k}
    \frac{\mathcal{C}_{j-k,n}(A_1,\ldots)\mathcal{C}_{k,s}(B_1,\ldots)q^{(2s+n)}(x_{t+1})}{2^s\Gamma(s+1)n!},
    \nonumber
  \end{eqnarray}
  where $B_j \equiv B_j(\{h_t^{(l)}\})$ appeared in Lemma
  \ref{lem:approx_mean_var}.

  (iii) Now we compare Eqs.\ (\ref{eq:true_predict}) and
  (\ref{eq:approx_predict}).  The coefficients, up to second order terms, in
  the former are
  \begin{eqnarray}
    C_1(x_{t+1}) =  \frac{q''(x_{t+1})}{2h_t''} - \frac{h_t'''q'(x_{t+1})}{2(h_t'')^2},
    \label{eq:C_1}
  \end{eqnarray}
  and
  \begin{eqnarray}
    C_2(x_{t+1}) &=&
    \frac{q^{(4)}(x_{t+1})}{8(h_t'')^2} 
    - \frac{5h_t'''q'''(x_{t+1})}{12(h_t'')^3}
    + \biggl[ \frac{5(h_t''')^2}{8(h_t'')^4}
      -  \frac{h_t^{(4)}}{4(h_t'')^3} \biggr]q''(x_{t+1}) \nonumber\\
    & & { } 
    +  \biggl[ \frac{2h_t'''h_t^{(4)}}{3(h_t'')^4}
      -  \frac{5(h_t''')^3}{8(h_t'')^5}
      - \frac{h_t^{(5)}}{8(h_t'')^3} \biggr]q'(x_{t+1}) .
  \end{eqnarray}

  For the first-order Laplace approximation ($\alpha=1$), the coefficient of
  order $\gamma^{-1}$ in Eq.\ (\ref{eq:approx_predict}) is
  \begin{equation}
    R_1(x_{t+1}) =  \frac{q''(x_{t+1})}{2h_t''},
  \end{equation}
  which does not correspond to $C_1(x_{t+1})$, and hence Eq.\
  (\ref{eq:approx_predict_t+1}) holds.

  For $\alpha \ge 2$, $R_1(x_{t+1})$ is as
  \begin{equation}
    R_1(x_{t+1}) =  \frac{q''(x_{t+1})}{2h_t''} - \frac{h_t'''q'(x_{t+1})}{2(h_t'')^2},
  \end{equation}
  which corresponds to $C_1(x_{t+1})$, and the first-order error term in Eq.\
  (\ref{eq:approx_predict_t+1}) is canceled.

  The second-order error term in Eq.\ (\ref{eq:approx_predict}) is calculated
  as
  \begin{equation}
    R_2(x_{t+1})=
    \frac{q^{(4)}(x_{t+1})}{8(h_t'')^2}  - \frac{h_t'''q'''(x_{t+1})}{4(h_t'')^3}
    + \biggl[ \frac{5(h_t''')^2}{8(h_t'')^4} - \frac{h_t^{(4)}}{4(h_t'')^3} \biggr]q''(x_{t+1}), 
  \end{equation}
  for $\alpha=2$, and
  \begin{eqnarray}
    R_2(x_{t+1}) &=&
    \frac{q^{(4)}(x_{t+1})}{8(h_t'')^2}  - \frac{h_t'''q'''(x_{t+1})}{4(h_t'')^3}
    + \biggl[ \frac{5(h_t''')^2}{8(h_t'')^4} - \frac{h_t^{(4)}}{4(h_t'')^3} \biggr]q''(x_{t+1})
    \nonumber\\
    & & { } +
    \biggl[\frac{2h'''h_t^{(4)}}{3(h_t'')^4} -\frac{5(h_t''')^3}{8(h_t'')^5} - \frac{h_t^{(5)}}{8(h_t'')^3}\biggr]
    q'(x_{t+1}),
    \label{eq:R_2}
  \end{eqnarray}
  for $\alpha \ge 3$.  Thus $R_2(x_{t+1}) \neq C_2(x_{t+1})$ and second-order
  error term in Eq.\ (\ref{eq:approx_predict_t+1}) remains for $\alpha \ge 2$.
  
  From (\ref{eq:C_1})--(\ref{eq:R_2}), the leading error term introduced by the
  Gaussian approximation is
\begin{equation}
\mathcal{E}^*_{t+1,1}(x_{t+1}) = R_1(x_{t+1})-C_1(x_{t+1}) =
 \frac{h_t'''q'(x_{t+1})}{2(h_t'')^2}
 \nonumber
\end{equation}
for $\alpha=1$, and
\begin{eqnarray}
\mathcal{E}^*_{t+1,2}(x_{t+1}) &=&
 R_2(x_{t+1})-C_2(x_{t+1}) \nonumber\\
&=&
    \frac{h_t'''q'''(x_{t+1})}{6(h_t'')^3}
    -  \biggl[ \frac{2h_t'''h_t^{(4)}}{3(h_t'')^4}
      -  \frac{5(h_t''')^3}{8(h_t'')^5}
      - \frac{h_t^{(5)}}{8(h_t'')^3} \biggr]q'(x_{t+1})
      \nonumber
\end{eqnarray}
for $\alpha=2$, and
\begin{equation}
\mathcal{E}^*_{t+1,2}(x_{t+1}) = 
 R_2(x_{t+1})-C_2(x_{t+1}) =
\frac{h_t'''q'''(x_{t+1})}{6(h_t'')^3}
\nonumber
\end{equation}
for $\alpha \ge 3$.
Thus if the condition (C.5) is satisfied, the leading error term
is bounded uniformly across time. 
We can confirm in the same way that the other error terms are also bounded uniformly.
  \hfill $\Box$
\end{Proof}

There are two sources of error in Eq.\ (\ref{eq:approx_predict_t+1}): first,
that due to the replacement of the true filtered distribution at time $t$ by a
Gaussian, $\sum_{j=\beta}^N\mathcal{E}^*_{t+1,j}(x_{t+1})\gamma^{-j}$, and,
second, that due to propagation from time $t$,
$\sum_{j=\nu}^{N-1}\mathcal{E}_{t+1,j+1}(x_{t+1})\gamma^{-j-1}$.  At each step,
the Gaussian approximation introduces an $O(\gamma^{-\beta})$ error into the
predictive distribution, where $\beta=1$ for $\alpha=1$ and $\beta=2$ for
$\alpha \ge 2$.  However, the errors propagated from the previous time-step
``move up'' one order of magnitude (power of $\gamma$).  Applying Eq.\
(\ref{eq:propagated_error}) repeatedly, we find that the leading error term,
$\mathcal{E}^*_{t,\beta}(x_{t})\gamma^{-\beta}$, which is generated at
time-step $t$, is propagated, by a strictly later time-step $u$, to be
$\mathcal{E}_{u,u-t+\beta}(x_u)\gamma^{-(u-t+\beta)}$ where
\begin{eqnarray}
\mathcal{E}_{u,u-t+\beta}(x_u) &=& 
q'(x_u) 
\prod_{k=t+1}^{u-1} \biggl[
\frac{1}{h_k''}\frac{\partial}{\partial x_k}
    \biggl( \frac{q'(x_k)}{p(x_k|y_{1:k-1})}  \biggr)
    \bigg|_{x_k=x_{k|k}}
\biggr]
\nonumber\\
& & { } \times
\biggl[
\frac{1}{h_{t}''}\frac{\partial}{\partial x_{t}}
    \biggl( \frac{\mathcal{E}^*_{t,\beta}(x_{t})}{p(x_{t}|y_{1:t-1})}  \biggr)
    \bigg|_{x_{t}=x_{t|t}}
\biggr] ~. \nonumber
\end{eqnarray}
The compounded error in time-step $u$ is then given by the summation of 
the propagated errors from $t=1$ to $u-1$ as
\begin{equation}
S_u = \sum_{t=1}^{u-1} \mathcal{E}_{u,u-t+\beta}(x_u)\gamma^{-(u-t+\beta)}
<  C^{-\beta}\sum_{t=1}^{u-1} (C\gamma^{-1})^{(u-t+\beta)}, \nonumber
\end{equation}
where the inequality holds under the condition (C.5), 
$C<\gamma$ is a constant which is independent of time $t$.
The right hand side in this equation converge on $O(\gamma^{-\beta-1})$ as 
$u\to\infty$, so that the compounded error after infinite time-step remains $O(\gamma^{-\beta-1})$.
The result is that the whole error term in the predictive distribution becomes
$O(\gamma^{-\beta})$, even if it started out smaller, but it does not grow
beyond that order.  
Theorem \ref{thm:predictive-dist} is then proved from Proposition \ref{thm:predict}
immediately:
\begin{Proof}[Theorem \ref{thm:predictive-dist}]
The LGFs start with an initial predictive distribution which does not involve 
any errors. Thus, from Proposition \ref{thm:predict} it is proved inductively that 
the error in the approximated predictive distribution is $O(\gamma^{-\beta})$ 
and uniformly bounded for $t \in \mathbb{N}$. 
\hfill $\Box$
\end{Proof}

\begin{Proof}[Theorem \ref{thm:filter}] (Sketch) 
Since the predictive distribution,
\begin{equation}
p(x_{t+1}|y_{1:t}) = \int p(x_{t+1}|x_t)p(x_t|y_{1:t})dx_t
\nonumber
\end{equation}
is the posterior expectation of $p(x_{t+1}|x_t)$ with respect to $x_t$, 
Theorem \ref{thm:filter} is proved in the same way as 
Theorem \ref{thm:predictive-dist} (replacing $p(x_{t+1}|x_t)$ by $g(x_t)$ in the proof of Theorem \ref{thm:predictive-dist}). 
\hfill $\Box$
\end{Proof}

\begin{Proof}[Theorem \ref{thm:stability}]
  From Proposition \ref{thm:predict}, the two predictive distributions at time $t$
  are given by
  \begin{equation}
    \hat{p}_1(x_t|y_{1:t-1}) =
    p(x_t|y_{1:t-1}) + \sum_{j=\nu}^{N} \mathcal{E}^{(1)}_{t,j}(x_t)\gamma^{-j} 
    + O(\gamma^{-N-1}), \nonumber
  \end{equation}
  and
  \begin{equation}
    \hat{p}_2(x_t|y_{1:t-1}) =
    p(x_t|y_{1:t-1}) + \sum_{j=\nu}^{N} \mathcal{E}^{(2)}_{t,j}(x_t)\gamma^{-j} 
    + O(\gamma^{-N-1}),  \nonumber
  \end{equation}
  where $\mathcal{E}^{(1)}_{t,j}(x_t) \neq \mathcal{E}^{(2)}_{t,j}(x_t)$.
  Applying the LGF to both predictive distributions introduces the same errors
  at time $t+1$, $\sum_{j=\beta}^N\mathcal{E}^*_{t+1,j}(x_{t+1})\gamma^{-j}$,
  which are canceled, while propagated errors from
  time-step $t$ to $t+1$ in both predictive distributions,
  $\sum_{j=\nu}^{N-1}\mathcal{E}^{(1)}_{t+1,j+1}(x_{t+1})\gamma^{-j-1}$ and
  $\sum_{j=\nu}^{N-1}\mathcal{E}^{(2)}_{t+1,j+1}(x_{t+1})\gamma^{-j-1}$ are not
  canceled. Then we get 
  $\hat{p}_1(x_{t+1}|y_{1:t}) -\hat{p}_2(x_{t+1}|y_{1:t}) = O(\gamma^{-\nu-1})$.
  Applying this procedure $u$ times completes the theorem.  \hfill $\Box$
\end{Proof}

\begin{Proof}[Theorem \ref{thm:smoothing}] Assume that the expectation at time
    $t+1$ satisfies
    \begin{equation}
      \hat{E}[g(x_{t+1})|y_{1:T}] = E[g(x_{t+1})|y_{1:T}] + O(\gamma^{-\beta}).
      \label{eq:expect_t+1}
    \end{equation}
    From Theorem \ref{thm:predictive-dist} and Eq.\ (\ref{eq:expect_t+1}), we
    obtain
    \begin{eqnarray}
      \int\frac{\hat{p}(x_{t+1}|y_{1:T})p(x_{t+1}|x_t)}{\hat{p}(x_{t+1}|y_{1:t})}dx_{t+1} 
      \nonumber
      &=&
      \int\frac{p(x_{t+1}|y_{1:T})p(x_{t+1}|x_t)}{p(x_{t+1}|y_{1:t})}dx_{t+1}+ O(\gamma^{-\beta}). \nonumber
    \end{eqnarray}
Using Theorem \ref{thm:filter}, the expectation at time $t$ is
    \begin{eqnarray}
      \hat{E}[g(x_t)|y_{1:T}] &=& 
      \int g(x_t) \hat{p}(x_t|y_{1:t}) \int\frac{p(x_{t+1}|y_{1:T})p(x_{t+1}|x_t)}{p(x_{t+1}|y_{1:t})}dx_{t+1} dx_t + O(\gamma^{-\beta}) \nonumber\\
      &=&
      \int g(x_t) p(x_t|y_{1:t}) \int\frac{p(x_{t+1}|y_{1:T})p(x_{t+1}|x_t)}{p(x_{t+1}|y_{1:t})}dx_{t+1} dx_t + O(\gamma^{-\beta}) \nonumber\\
      &=& E[g(x_t)|y_{1:T}]  + O(\gamma^{-\beta}). \nonumber
    \end{eqnarray}
    The initial smoothed distribution of the backward recursion is given by the
    filtered distribution $\hat{p}(x_T|y_{1:T})$, which satisfies
    $\hat{E}[x_T|y_{1:T}]=E[x_T|y_{1:T}]+O(\gamma^{-\beta})$ by
    theorem~\ref{thm:filter}. 
    Then, the theorem is proved inductively.
     \hfill $\Box$
\end{Proof}

\section{Numerical Computation for second derivatives}
\label{appendix:second-derivs}
We describe the numerical algorithm for computing the Hessian matrix, as
promised in Section \ref{sec:implementation}.

The Laplace approximation requires the second derivative (or the Hessian
matrix) of the log-likelihood function evaluated at its maximum.  However, it
is often difficult, and even more often tedious, to get correct analytical
derivatives of the log-likelihood function.  In such cases accurate numerical
computations of the derivative may be used, as follows.  Consider calculating
the second derivative of $l(x)$ at $x_0$ for the one-dimensional case.  For
$n=0,1,2,\ldots$ and $c>1$, define the second central difference quotient,
\begin{equation}
A_{n,0} = [l(x_0+c^{-n}h_0) + l(x_0-c^{-n}h_0) - 2l(x_0)]/(c^{-n}h_0)^2,
\nonumber
\end{equation}
and then for $k=1,2,\ldots,n$ compute
\begin{equation}
A_{n,k} = A_{n,k-1} + \frac{A_{n,k-1}-A_{n-1,k-1}}{c^{2(k+1)}-1}.
\label{eq:2nd-central-diff}
\end{equation}
When the value of $|A_{n,k}-A_{n-1,k}|$ is sufficiently small, $A_{n,k+1}$ is
used for the second derivative.

This algorithm is an iterated version of the second central difference formula,
often called {\it Richardson extrapolation}, producing an approximation with an
error of order $O(h^{2(k+1)})$~\cite{Dahlquist-Bjorck-NumericalMethods}.

In the $d$-dimensional case of a second-derivative approximation at a maximum,
\citeasnoun{Kass-computing-observed-information} proposed an efficient numerical
routine which reduces the computation of the Hessian matrix to a series of
one-dimensional second-derivative calculations.  The trick is to apply the
second-difference quotient to suitably-defined functions $f$ of a single
variable $s$ as follows.

\begin{enumerate}
\item Initialize the increment $\bm{h}=(h_1,\ldots,h_d)$.
\item Find the maximum of $l(\bm{x})$, and call it $\hat{\bm{x}}$.
\item Get all unmixed second derivatives for each $i=1$ to $d$, using the
  function
  \begin{eqnarray}
    x_i &=& \hat{x}_i + s \nonumber\\ 
    x_j &=& \hat{x}_j \quad \mathrm{for}\ j\ \mathrm{not\ equal\ to}\ i \nonumber\\
    f(s) &=& l(\bm{x}(s)). \nonumber
  \end{eqnarray}
  Compute the second difference quotient; then repeat and extrapolate until the
  difference in successive approximations meets a relative error criterion, as
  in (\ref{eq:2nd-central-diff}); store as diagonal elements of the Hessian
  matrix array, $l''_{i,i}=f''(0)$.
\item Similarly, get all the mixed second derivatives. For each $i=1$ to $d$,
  for each $j=i+1$ to $d$, using the function
  \begin{eqnarray}
    x_i &=& \hat{x}_i + s/\sqrt{l''_{i,i}} \nonumber\\ 
    x_j &=& \hat{x}_j + s/\sqrt{l''_{j,j}} \nonumber\\ 
    x_k &=& \hat{x}_k \quad \mathrm{for}\ k\ \mathrm{not\ equal\ to}\ i\ \mathrm{or}\ j \nonumber\\
    f(s) &=& l(\bm{x}(s)). \nonumber
  \end{eqnarray}
  Compute the second difference quotient; then repeat and extrapolate until
  difference in successive approximations is less than relative error criterion
  as in (\ref{eq:2nd-central-diff}); store as off-diagonal elements of the
  Hessian matrix array, $l''_{i,i}=(f''(0)/2-1)\sqrt{l''_{i,i}l''_{j,j}}$.
\end{enumerate}

In practice, the increment 
for computing the Hessian at time $t$ would be taken to be 
$h_i = 0.1 \times \sqrt{v_{t|t-1}^{(i,i)}}$, $i=1,2,\ldots,d$, 
where $v_{t|t-1}^{(i,i)}$ is the $(i,i)$-element of the covariance matrix of 
the predictive distribution at time $t$.

\section{Laplace's Method}\label{app:laplace}

Here, we briefly describe Laplace's method, especially the details used in the
proofs of Lemma 6 and Proposition 7.

We consider the following integral,
\begin{equation}
I(\gamma) = \int g(x) e^{-\gamma h(x)} dx,
\label{eq:laplace_form}
\end{equation}
where $x\in\mathbb{R}$; $\gamma$, the expansion parameter, is a large positive
real number; $h(x)$ and $g(x)$ are independent of $\gamma$ (or very weakly
dependent on $\gamma$); and the interval of integration can be finite or
infinite.  Laplace's method approximates $I(\gamma)$ as a series expansion in
descending powers of $\gamma$.  There is a computationally efficient method to
compute the coefficients in this infinite asymptotic expansion (Theorem 1.1 in
\cite{Wojdylo}).  Suppose that $h(x)$ has an interior minimum at $x_0$, and
$h(x)$ and $g(x)$ are assumed to be expandable in a neighborhood of $x_0$ in
series of ascending powers of $x$. Thus, as $x \to x_0$,
\begin{equation}
  h(x) \sim h(x_0) + \sum_{s=0}^{\infty}a_s(x-x_0)^{s+2},
\nonumber
\end{equation}
and
\begin{equation}
  g(x) \sim \sum_{s=0}^{\infty}b_s(x-x_0)^s,
  \nonumber
\end{equation}
in which $a_0,b_0 \neq 0$.

Let us introduce two dimensionless sets of quantities, $A_i \equiv a_i/a_0$ and
$B_i \equiv b_i/b_0$, as well as the constants $\alpha_1 = 1/a_0^{1/2}$ and
$c_0=b_0/a_0^{1/2}$.  Then the integral in \ref{eq:laplace_form}) can be
asymptotically expanded as
\begin{equation}
  I(\gamma) \sim c_0e^{-\gamma h(x_0)} \sum_{s=0}^{\infty}
  \Gamma(s+\frac{1}{2})\alpha_1^{2s}c^*_{2s}\gamma^{-s-\frac{1}{2}},
  \nonumber
\end{equation}
where 
\begin{equation}
  c^*_s = \sum_{i=0}^s B_{s-i} \sum_{j=0}^i{\left(
    \begin{array}{cc}
      -\frac{s+1}{2} \\ j
    \end{array}
    \right)
    \mathcal{C}_{i,j}(A_1,\ldots)} ~, \nonumber
\end{equation}
where $\mathcal{C}_{i,j}(A_1,\ldots)$ is a partial ordinary Bell polynomial,
the coefficient of $x^i$ in the formal expansion of $(A_1x+A_2x^2+\cdots)^j$.
$\mathcal{C}_{i,j}(A_1,\ldots)$ can be computed by the following recursive
formula,
\begin{equation}
  \mathcal{C}_{i,j}(A_1,\ldots)=\sum_{m=j-1}^{i-1}{A_{i-m}\mathcal{C}_{m,j-1}(A_1,\ldots)}~, \nonumber
\end{equation}
for $1 \ge j \ge i$.  Note that $\mathcal{C}_{0,0}(A_1,\ldots)=1$, and
$\mathcal{C}_{i,0}(A_1,\ldots)= \mathcal{C}_{0,j}(A_1,\ldots)=0$ for all
$i,j>0$.

\section{The Population Vector Algorithm} \label{appendix:pva}

The {\em population vector algorithm} (PVA) is a standard method for neural
decoding, especially for directionally-sensitive neurons like the
motor-cortical cells recorded from in the experiments we analyze \cite[pp.\
97--101]{Dayan-Abbott}.  Briefly, the idea is that each neuron $i$, $1 \leq i
\leq N$, has a preferred motion vector $\bm{\theta}_i$, and the expected
spiking rate $\lambda_i$ varies with the inner product between this vector and
the actual motion vector $\bm{x}(t)$,
\begin{equation}
  \label{eqn:inner-product-firing-rate} \frac{\lambda_i(t) - r_i}{\Lambda_i} = \bm{x}(t) \cdot \bm{\theta}_i ~,
\end{equation}
where $r_i$ is a baseline firing rate for neuron $i$, and $\Lambda_i$ a maximum
firing rate.  ((\ref{eqn:inner-product-firing-rate}) corresponds to a cosine
tuning curve.)  If one observes $y_i(t)$, the actual neuronal counts over some
time-window $\Delta$, then averaging over neurons and inverting gives the {\em
  population vector}
\begin{equation}
\bm{x}_{\mathrm{pop}}(t) = \sum_{i=1}^{N}{\frac{y(t) - r_i \Delta}{\Lambda_i \Delta} \bm{\theta}_i} ~,
\nonumber
\end{equation}
which the PVA uses as an estimate of $\bm{x}(t)$.  If preferred vectors
$\bm{\theta}_i$ are uniformly distributed, then $\bm{x}_{\mathrm{pop}}$
converges on a vector parallel to $\bm{x}$ as $N \rightarrow \infty$, and is in
that sense unbiased \cite[p.\ 101]{Dayan-Abbott}.  If preferred vectors are not
uniform, however, then in general the population vector gives a biased
estimate.

\section{Real data analysis}
\label{sec:real-data-trajectory}
\begin{figure}[htbp]
\begin{center}
\resizebox{!}{0.86\textheight}{\includegraphics{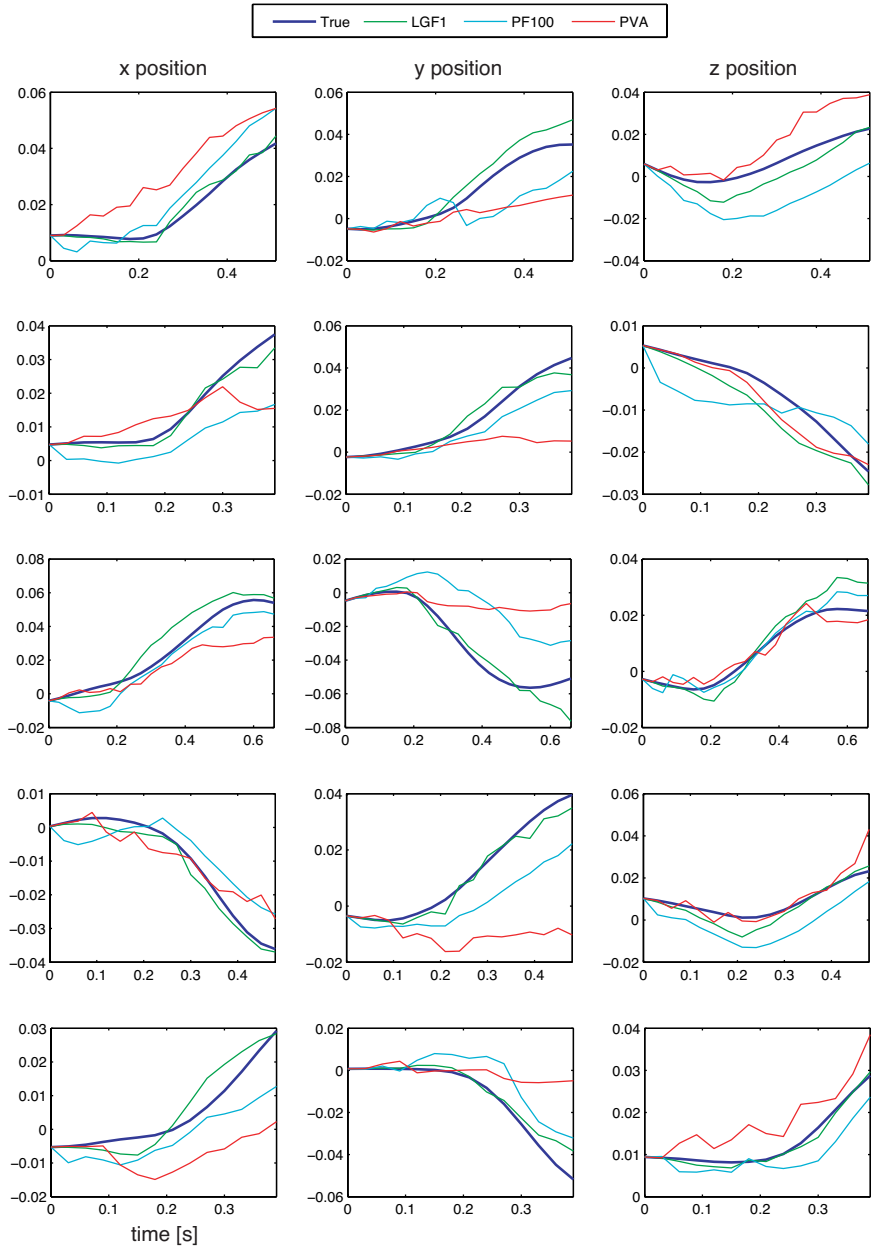}}
\end{center}
\caption{The trajectories of the cursor position.  ``True'': actual
  trajectory. ``LGF1'': trajectories estimated by first-order LGF,
  respectively. ``PF100'': trajectory estimated by the particle filter with 100
  particles. ``PVA'': trajectory estimated by the population vector algorithm.
  The trajectories estimated by the LGF2 are not shown; they are similar to
  those estimated by the LGF1.  }
\label{fig:real-data-trajectory} 
\end{figure}

Figure \ref{fig:real-data-trajectory} shows trajectories of the true and
estimated (by LGF, PF-100 and PVA) cursor position of the real data analysis.
It is seen that the LGF provides better estimation than either the PF-100 or
the PVA.

\clearpage

\bibliographystyle{ECA_jasa}
\bibliography{mybibliography}

\end{document}